\documentclass[twocolumn]{aastex631}
\usepackage{multirow}
\usepackage{color}
\usepackage{umoline}

\received{}
\revised{}
\accepted{June 17, 2021}
\submitjournal{AJ}

\shortauthors{Hong et al.}

\begin{document}
\title{New sub-grouping of multiple stellar populations in NGC 2808 based on low-resolution spectroscopy}

\author[0000-0003-4364-6744]{Seungsoo Hong}
\affiliation{Center for Galaxy Evolution Research \& Department of Astronomy, Yonsei University, Seoul 03722, Republic of Korea}

\author[0000-0001-7277-7175]{Dongwook Lim}
\affiliation{Zentrum f\"ur Astronomie der Universit\"at Heidelberg, Astronomisches Rechen-Institut, M\"onchhofstr. 12-14, 69120 Heidelberg, Germany}

\author[0000-0001-6812-4542]{Chul Chung}
\affiliation{Center for Galaxy Evolution Research \& Department of Astronomy, Yonsei University, Seoul 03722, Republic of Korea}

\author[0000-0002-0432-6847]{Jaeyeon Kim}
\affiliation{Zentrum f\"ur Astronomie der Universit\"at Heidelberg, Astronomisches Rechen-Institut, M\"onchhofstr. 12-14, 69120 Heidelberg, Germany}

\author{Sang-Il Han}
\affiliation{Department of Science Education, Ewha Womans University, Seoul 03760, Republic of Korea}

\author[0000-0002-2210-1238]{Young-Wook Lee}
\affiliation{Center for Galaxy Evolution Research \& Department of Astronomy, Yonsei University, Seoul 03722, Republic of Korea}

\correspondingauthor{Young-Wook Lee} 
\email{ywlee2@yonsei.ac.kr}

\begin{abstract}
We performed low-resolution spectroscopy for the red giant branch stars in an intriguing globular cluster (GC) NGC 2808, which hosts subpopulations with extreme helium and light-element abundances. In order to trace N, C, and Ca abundance differences among subpopulations, we measured CN, CH, and Ca II H\&K spectral indices, respectively. We identified four subpopulations (G1, G2, G3, and G4) from CN and CH strength, with CN-weak/CH-strong G1, CN-intermediate/CH-strong G2, CN-strong/CH-intermediate G3, and CN-strong/CH-weak G4. Compared to [Na/O] from high-resolution spectroscopy, we show that CN index can more clearly separate G1 and G2. Since CN traces N abundance in a GC, it implies that G1 and G2 would show a larger difference in [N/Fe] compared to [Na/Fe], as predicted by chemical evolution models. Later generation stars G3 and G4, however, are better separated with high-resolution spectroscopy. We also found that G4 shows a stronger Ca II H\&K line strength compared to that of G1, but we suspect this to be a result of unusually strong He enhancement and/or Mg depletion in G4 of this GC. This work illustrates that combining low- and high-resolution spectroscopic studies can improve the separation of subpopulations in GCs.

\end{abstract}


\section{Introduction} \label{section:intro}
Numerous and dedicated observations made during the last two decades have established that most globular clusters (GCs) host multiple stellar populations (\citealt{Lee99,Gra12,Ren15,BL18}, and references therein). They usually show abundance variations in light-elements with little or no dispersion in heavy-elements (see, e.g., \citealt{Carr09a}). The abundance dispersions in such ``normal'' GCs are generally explained as a self-enrichment by asymptotic giant branch (AGB) stars (\citealt{DC04,Der10,Dan16}), fast-rotating massive stars (FRMSs; \citealt{Dec07a,Dec07b,Kra13}), massive interacting binaries (MIBs; \citealt{DeMink09,Bas13}), or supermassive stars (SMSs; \citealt{DH14,Gie18}). A few GCs showing abundance variations in heavy-elements, however, have been discovered, such as $\omega$ Centauri and M22 (\citealt{Lee99,jwlee09,Mar09,JP10}). Because heavy-elements dispersions indicate an enrichment by supernovae (SNe), such ``peculiar'' GCs would have been more massive in the past to retain SNe ejecta (\citealt{Timm95,Baum08,ST17}). 

Among the ``normal'' GCs, NGC 2808 is one of the most studied GCs due to its intriguing features on the color-magnitude diagram (CMD). The red giant branch (RGB) does not show the discreteness on the optical or near-infrared CMD as other GCs without significant spread in heavy-elements. Multimodalities, however, are observed along both the main sequence (MS) (\citealt{Pio07}) and the horizontal branch (HB) including the extreme blue HB (see, e.g., \citealt{Dal11}). From these features, the presence of helium-enhanced multiple populations has been postulated (\citealt{Lee05,Pio07}), and the helium variation has indeed been confirmed by spectroscopy (\citealt{Pas11,Mar14}). In addition, \textit{Hubble Space Telescope} (HST) ultraviolet photometry and high-resolution spectroscopy have revealed the presence of at least 5 subpopulations with different light-element abundances (\citealt{Mil15,Carr15}). Note that a GC with more than 5 subpopulations is a rare case, except for $\omega$ Centauri and NGC 7089 (see, e.g., \citealt{Mil17}). In spite of these extreme features, NGC 2808 has been considered as a ``normal'' GC because its stars do not show clear signs of heavy elements spreads, which is also supported by the recent low-resolution spectroscopic study using MUSE (\citealt{Lat19}). \citet{Hus20}, using the same spectra, however, reported an unexpected Ca II triplet metallicity variation in this GC, suggesting a possibility of a little spread in heavy elements. 

In order to further investigate light and heavy elements abundance patterns in NGC 2808, in this paper, we measured spectral indices for CN, CH, and Ca II H\&K lines as well known tracers of N, C, and Ca abundances, respectively, for the sample of 91 RGB stars in this GC. These indices are widely used in the multiple population studies (\citealt{NS83,SS91,Smi96,Coh99,Har03,Pan10,Smol11,Simp17,Ger20}, and references therein), including our previous works (\citealt{Han15,Lim15,Lim16,Lim17}).

\begin{figure}
	\centering{\includegraphics[scale=0.5]{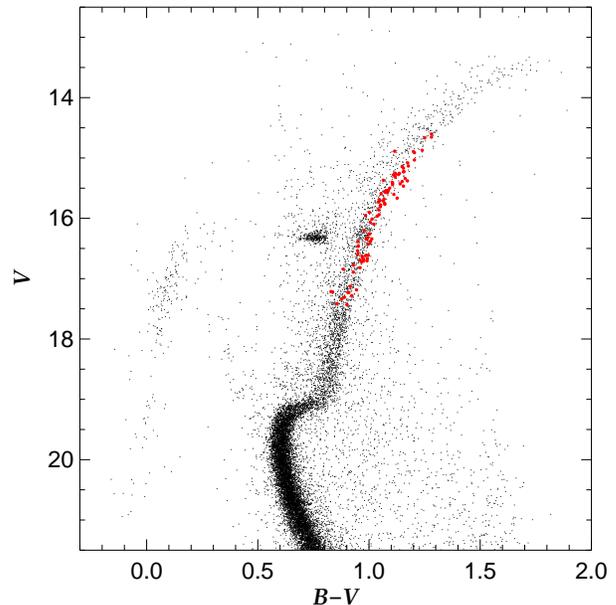}}
	\caption{The CMD for NGC 2808 in the (\textit{V, B-V}) plane obtained at the CTIO 4m Blanco telescope. The red circles indicate final samples used in this analysis.}
	\label{fig_cmd}
\end{figure}

\begin{figure*}
	\centering{\includegraphics[scale=0.85]{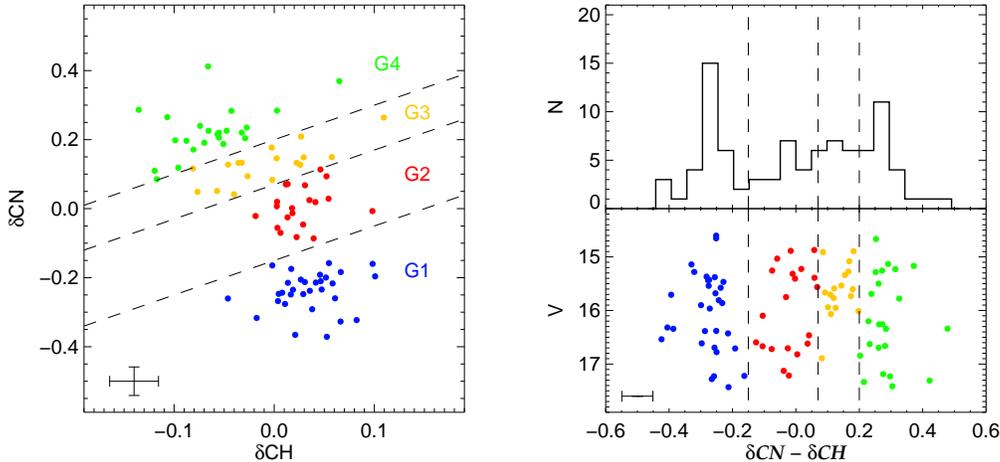}}
	\caption{Left: Distribution of our sample stars on the $\delta$CN-$\delta$CH plane.	We have divided 4 subpopulations on this plane, G1 (blue; CN-weak/CH-strong), G2 (red; CN-intermediate/CH-strong), G3 (orange; CN-strong/CH-intermediate), and G4 (green; CN-strong/CH-weak). Dashed lines indicate the $\delta CN-\delta CH$ to be -0.15, 0.07, and 0.20, respectively. Right: Histogram and magnitude distribution of $\delta CN-\delta CH$. The four peaks, each of which corresponds to G1, G2, G3, and G4, are identified in the histogram.}
	\label{fig_cnch}
\end{figure*}

\section{Observations and Data Reduction} \label{section:obs}
The spectroscopic observations were performed with the Wide Field Reimaging CCD Camera (WFCCD) mounted on the du Pont 2.5m telescope at Las Campanas Observatory (LCO). The spectroscopic target stars were selected from the photometric data obtained at the Cerro Tololo Inter-American Observatory (CTIO) 4m Blanco telescope and reduced as described in \citet{Han09}. For the multi-object spectroscopy, a total of 7 multi-slit masks were designed, each of them contains $\sim$21 slits of 1{\arcsec}.2 width. The observations were made in Febuary 2012 and April 2016. The HK grism, which provides a central wavelength of 3700 {\AA} with a dispersion of 0.8 {\AA}/pixel, was used in order to securely detect CN, CH, and Ca II H\&K lines near 4000 {\AA}. At least two science images, together with three flats and an arc lamp frame for calibration, were obtained for each mask. Exposure times for science frames are longer than 1500s depending on the brightness grouping of stars. We used the modified version of the WFCCD reduction package and IRAF\footnote{IRAF is distributed by the National Optical Astronomy Observatory, which is operated by the Association of Universities for Research in Astronomy (AURA) under a cooperative agreement with the National Science Foundation.} for the data reduction and radial velocity (RV) measurements as employed in previous studies (\citealt{Pro06,Lim15,Lim16,Lim17}). The RV distribution of our sample stars is roughly gaussian with a peak value of $\sim$92 km/s, which is comparable to but somewhat smaller than the value of $\sim$101 km/s listed in \citet{Harris96}. This is probably because the typical uncertainty of RV values from our low-resolution measurement appears to be relatively large ($\sim$30 km/s) compared to those from high-resolution spectroscopy. After the rejection of low signal-to-noise (S/N) ratio samples ($<$ 8 at 3900 {\AA}) and non-member stars (RV $>$ 2.5 $\sigma$ of the mean velocity of the GC), 91 stars were finally used for our analysis, which are identified in Figure~\ref{fig_cmd}. The typical S/N ratio for the selected sample stars is $\sim$17 at 3900 {\AA}.

Following \citet{Har03} and \citet{Lim15}, we calculated the S(3839), CH(4300), and HK' indices to measure the strengths of CN, CH, and Ca II H\&K lines of each target stars in NGC 2808, respectively. The definitions for these indices are	
\begin{eqnarray*}
{\rm CN}(3839) & = & -2.5 \log{\frac{F_{3861-3884}}{F_{3894-3910}}} , \\
{\rm CH}(4300) & = & -2.5 \log{\frac{F_{4285-4315}}{0.5F_{4240-4280}+0.5F_{4390-4460}}} , \\
{\rm HK'}  & = & -2.5 \log{\frac{F_{3916-3985}}{2F_{3894-3911}+F_{3990-4025}}},			
\end{eqnarray*}
where $F_{3861-3884}$, for example, is the integrated flux from 3861 to 3884 {\AA}. The measurement error of each index was estimated from Poisson statistics in the flux measurements (\citealt{VE06}). We also measured $\delta$-indices ($\delta$CN, $\delta$CH, and $\delta$HK$'$) as the difference between the original index and the least-square fitting lines of the full sample in this GC, in order to compare chemical abundances of stars excluding the effects of magnitude (gravity and temperature). The measured indices and the errors are listed in Table~\ref{tab_data}.

\section{Results of Low-Resolution Spectroscopy} \label{section:result_lrs}
\begin{figure}
	\centering{\includegraphics[scale=0.5]{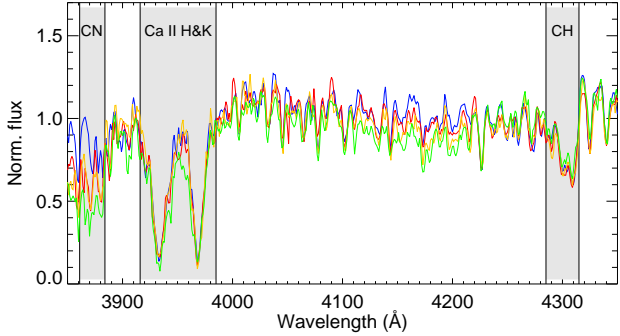}}
	\caption{Example of the continuum-normalized spectra for the stars belonging to G1 (Blue; V=15.4), G2 (Red; V=15.4), G3 (orange; V=15.6), and G4 (green; V=15.2). The CN, CH, and Ca II H\&K bands are indicated in grey.}
	\label{fig_spectra}
\end{figure}

We first define subpopulations based on the CN-CH diagram. For the RGB stars in NGC 2808, the $\delta$CN and $\delta$CH indices are anti-correlated as shown in the left panel of Figure~\ref{fig_cnch}. This CN-CH anti-correlation is one of the well-known features in low-resolution spectroscopic studies for ``normal'' GCs (\citealt{SS91,Smi96,Coh99,Har03,Pan10,Smol11,Lim15}, and references therein). NGC 2808 follows this general trend as expected from the homogeneity in iron abundance among its member stars (\citealt{Gra11,Carr15}). High-resolution spectroscopic studies found a similar anti-correlation between Na and O abundances for ``normal'' GCs, which is widely used to define subpopulations in GCs. For instance, \citet{Carr09a} defined P (Na-poor/O-rich stars), I (Na-/O-intermediate stars), and E (Na-rich/O-poor stars) subpopulations on the Na-O plane (see also \citealt{Carr14}). In a similar manner, we have divided 4 subpopulations along the CN-CH sequence, CN-weak/CH-strong G1, CN-intermediate/CH-strong G2, CN-strong/CH-intermediate G3\footnote{Referee has pointed out that stars in G3 appears to be restricted among the brightest ones, apart from one star. It is interesting to note that the V magnitude of the RGB-bump in NGC 2808 is 16.15 $\pm$ 0.05 (\citealt{Fer99}), and almost all the stars in G3 are brighter than this luminosity. It is also possible that our G3 is in fact a combination of the tails of the distributions of G2 and of G4, although the histogram in the right panel of Figure~\ref{fig_cnch} suggests that it is more likely a separate population.}, and CN-strong/CH-weak G4. These four subpopulations are indentified in the $\delta CN - \delta CH$ histogram as shown in the right panel of Figure~\ref{fig_cnch}. Qualitatively, these four peaks are consistent with those reported by \citet{N87} from CN photometry. Our spectroscopic study is based on $\delta CN - \delta CH$, and, therefore, the G3 and G4 subpopulations are more clearly separated. Examples of observed spectra for each subpopulation are shown in Figure~\ref{fig_spectra}.

\begin{figure}
	\centering{\includegraphics[scale=0.6]{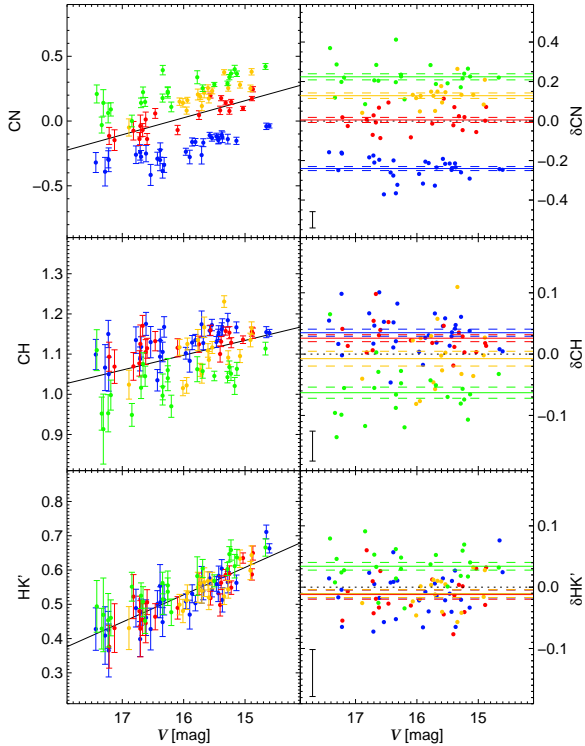}}
	\caption{Left panels: measured spectral indices (CN, CH, HK') as functions of \textit{V} magnitude for RGB stars in NGC 2808.
		The four subpopulations defined in Figure~\ref{fig_cnch} are indicated with the same color coding.
		Right panels: the $\delta$CN, $\delta$CH, and $\delta$HK' indices plotted against \textit{V} magnitude. The mean value and the error of the mean for each subpopulation are denoted by solid and dashed lines, respectively. The vertical bars in the lower left corner indicate the typical measurement error for each index.}
	\label{fig_indices}
\end{figure}

\begin{deluxetable}{ccccc}
	\tablecaption{The average values of $\delta$CN, $\delta$CH, and $\delta$HK' indices in all sub-groups.} \label{tab_index}
	\tablehead{
		\colhead{Subpop.} & \colhead{Mean} & \colhead{Error$_{mean}$} & \colhead{Stddev.} & \colhead{N}
	}
	\startdata
	& & $\delta$CN & & \\
	G1 & -0.2405 &  0.0102 &  0.0566 & 31 \\
	G2 &  0.0055 &  0.0129 &  0.0579 & 20 \\
	G3 &  0.1284 &  0.0138 &  0.0568 & 17 \\
	G4 &  0.2240 &  0.0154 &  0.0740 & 23 \\
	\hline
	& & $\delta$CH & & \\
	G1 &  0.0349 &  0.0058 &  0.0321 & 31 \\
	G2 &  0.0260 &  0.0057 &  0.0254 & 20 \\
	G3 & -0.0074 &  0.0121 &  0.0498 & 17 \\
	G4 & -0.0629 &  0.0091 &  0.0436 & 23 \\
	\hline
	& & $\delta$HK' & & \\
	G1 & -0.0110 &  0.0064 &  0.0358 & 31 \\
	G2 & -0.0120 &  0.0076 &  0.0341 & 20 \\
	G3 & -0.0106 &  0.0064 &  0.0262 & 17 \\
	G4 &  0.0341 &  0.0063 &  0.0303 & 23 \\
	\enddata
\end{deluxetable}

Figure~\ref{fig_indices} shows the measured spectral indices (CN, CH, and HK') as functions of V magnitude obtained from our CTIO photometry. The strengths of these indices increase as V magnitude decreases, which is due to the effects of temperature and surface gravity. In order to compare chemical abundances of stars without the magnitude effects, we calculated the differences in mean values of the $\delta$-indices between subpopulations, which are summarized in Table~\ref{tab_index}. The top-right and middle-right panels of Figure~\ref{fig_indices} show that the $\delta$CN and $\delta$CH indices have significant spreads that are clearly larger than typical measurement errors. The $\delta$CN enhancements of G2, G3, and G4 compared to G1 are 0.25, 0.4, and 0.45, respectively. Unlike $\delta$CN index, $\delta$CH decreases towards the later generations. The mean values of G3 and G4 are decreased by 0.045 and 0.095 compared to G1, respectively, while the difference in $\delta$CH between G1 and G2 is negligible. The $\delta$CN and $\delta$CH indices, in other words, are not linearly anti-correlated, but the relation between them has a ``banana'' shape (see the left panel of Figure~\ref{fig_cnch}), like the well-known Na-O anti-correlation. When the G2, G3, and G4 are combined to G2+ as in our previous works\footnote{\citet{Lim15,Lim16} defined G1 and G2+ in a GC from the two distinct RGBs in CMD obtained from the narrow-band photometry with Ca+CN filter. From the follow-up low-resolution spectroscopy, however, they showed that G1 and G2+ correspond to CN-weak and CN-strong groups, respectively.}, the differences in $\delta$CN and $\delta$CH indices between G1 and G2+ are $\sim$0.36 and $\sim$0.05, respectively. This difference in $\delta$CN is somewhat larger compared to other GCs showing CN-CH anti-correlation. For instance, G1 and G2+ in both NGC 288 and NGC 362 are separated by $\sim$0.31 in $\delta$CN. When the G4 is excluded, the difference between G1 and `G2+G3' in NGC 2808 becomes more comparable to other GCs (from $\sim$0.36 to $\sim$0.32). Interestingly, the G4 of NGC 2808 shows a `jump' in $\delta$HK' index, while G2 and G3 show no appreciable separations from G1. The difference in $\delta$HK' index between G1 and G4 is 0.045 (see the bottom right panel of Figure~\ref{fig_indices}), which is significant at 5.0$\sigma$ level. The abundance pattern that G4 shows (i.e. $\delta$CN increased, $\delta$CH decreased, $\delta$HK' increased) is very unusual in the sense that most CN-enhanced and CH-depleted subpopulation in a ``normal'' GC does not show an enhancement in $\delta$HK' index. In ``peculiar'' GCs, on the other hand, the metal-enriched later generation, traced by $\delta$HK' index, shows enhancements in both $\delta$CN and $\delta$CH indices (\citealt{Han15,Lim15,Lim17}). We note that subpopulations showing a similar abundance pattern (Na increased, O decreased, Ca increased) have been reported by \citet{CB21} in 8 GCs, although the definition of Ca excess in that study is the excess of [Ca/H] at given [Mg/H] with respect to field stars. We also note that the spread in $\delta$HK' index, caused by G4, in NGC 2808 is qualitatively consistent with the result of the recent spectroscopic study by \citet{Hus20}, which reported an unexpected variation of Ca II triplet line strengths in this GC. In Section~\ref{subsection:Ca_spread}, we discuss the probable cause of the spread in Ca line strengths observed in this GC.

\begin{figure}
	\centering{\includegraphics[scale=0.5]{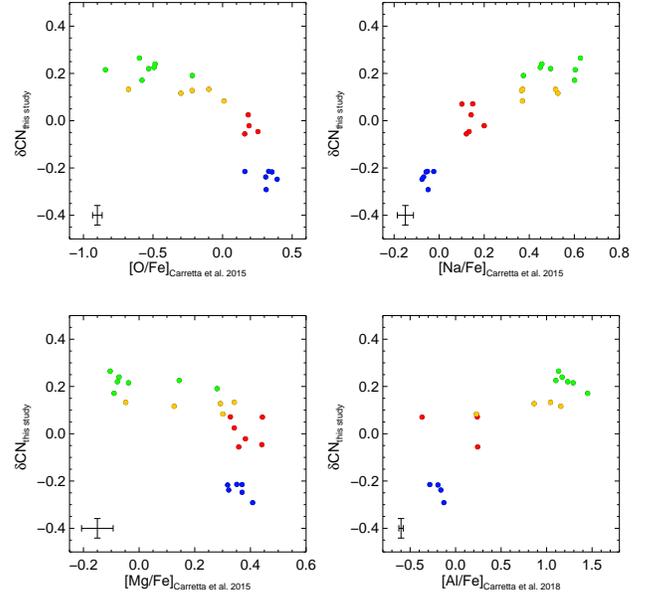}}
	\caption{$\delta$CN index as functions of O, Na, Mg, and Al abundances for RGB stars of NGC 2808. The same color coding defined in Figure~\ref{fig_cnch} is used. In each panel, typical measurement errors are shown in the lower left corner.}
	\label{fig_cn_hrs}
\end{figure}

\begin{figure}
	\centering{\includegraphics[scale=0.5]{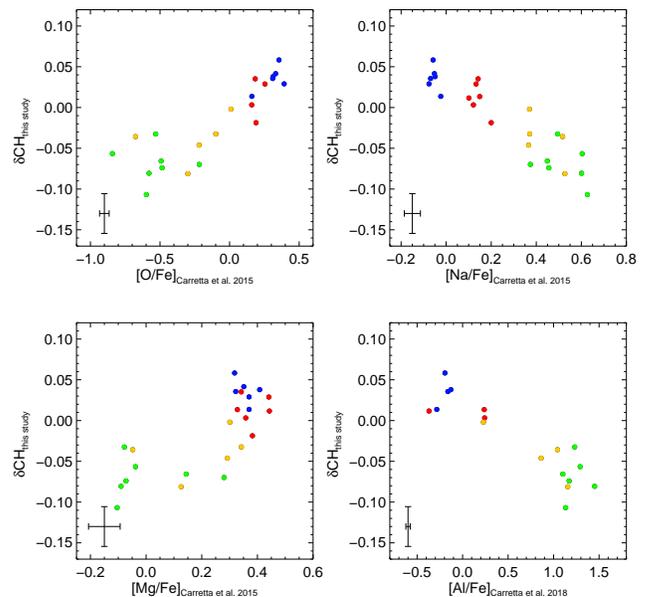}}
	\caption{Same as Figure~\ref{fig_cn_hrs}, but for $\delta$CH index.}
	\label{fig_ch_hrs}
\end{figure}

\begin{figure*}
	\centering{\includegraphics[scale=1.0]{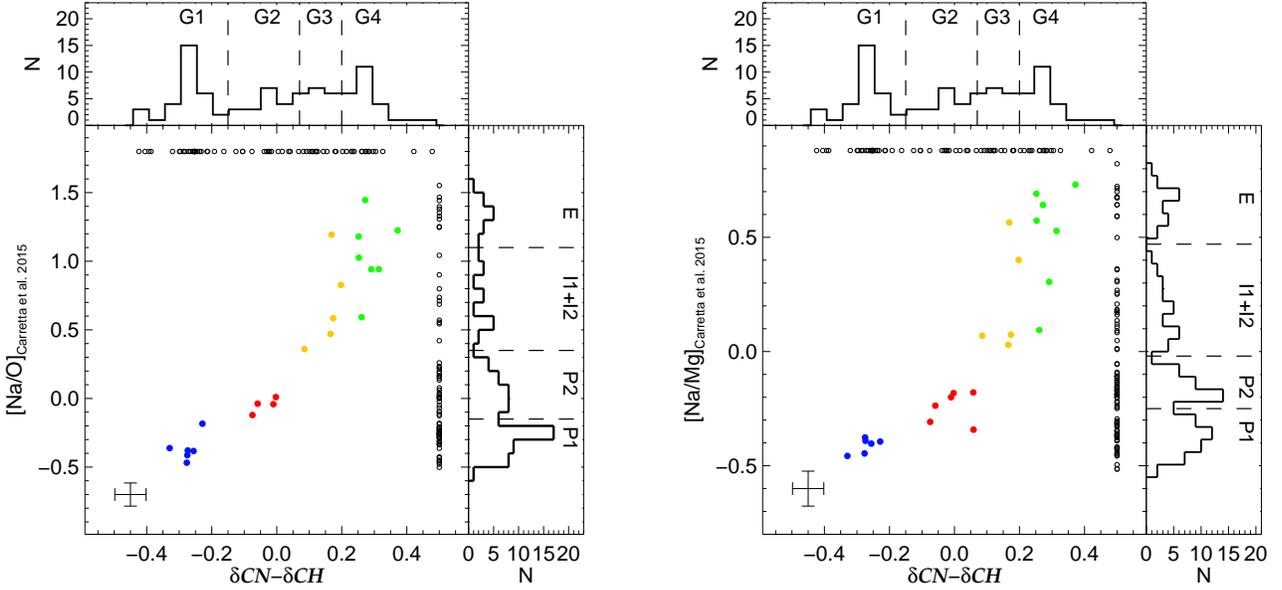}}
	\caption{Left: Relation between $\delta CN - \delta CH$ and [Na/O] for the sample stars common with \citet{Carr15}. The black open circles are the remaining unmatched stars. The color coding is the same as in previous figures. Histograms, which include unmatched stars, for $\delta CN - \delta CH$ and [Na/O] are also shown at each axis. Right: Same as left panel, but for [Na/Mg]. Note that our G1, G2, G3, and G4 are well mathced with the P1, P2, I1+I2, and E of \citet{Carr15}, respectively, but G1 and G2 are more clearly separated with $\delta CN - \delta CH$.}
	\label{fig_comp}
\end{figure*}

\begin{figure}
	\centering{\includegraphics[scale=0.5]{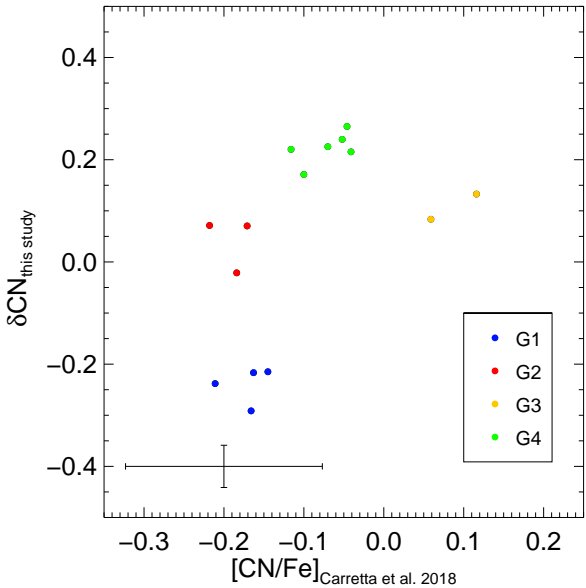}}
	\caption{Comparison of our $\delta$CN with [CN/Fe] obtained from high-resolution spectroscopy (\citealt{Carr18}). Different colors indicate the four subpopulations defined in Figure~\ref{fig_cnch}. Compared to the [CN/Fe], our $\delta$CN appears to be more sensitive to the abundance variation between G1 and G2.}
	\label{fig_cn_cnfe}
\end{figure}

\section{Comparision with High-Resolution Spectroscopy} \label{section:comp_hrs}
A high-resolution spectroscopic survey with FLAMES/VLT provides high-quality and homogeneous spectroscopic data set for the elemental abundances of 140 RGB stars in NGC 2808 (\citealt{Carr03,Carr04,Carr06a,Carr18,Carr06b,Carr14,Carr15}; see also \citealt{Gra11,Mar14}). We cross-matched our low-resolution data with this high-resolution data set, which results in the common sample of 24 stars. Figure~\ref{fig_cn_hrs} (Figure~\ref{fig_ch_hrs}) shows the relations, for the common sample stars, between $\delta$CN ($\delta$CH) and abundances of key elements in the multiple population studies, where the abundance ratios are adopted from \citet{Carr15} and \citet{Carr18}. The $\delta$CN is anti-correlated with O and Mg, while it is correlated with Na and Al. The $\delta$CH shows the opposite patterns, i.e., anti-correlated with Na and Al, and correlated with O and Mg. These trends are in good agreements with previous studies for GCs (\citealt{Sne92,Smi96,Mar08,Lim16,Lim17}). In particular, \citet{Smi13} and \citet{Smi15} noticed a non-linear relation between CN and Na (and O). The relation between $\delta$CN and Na, together with that between $\delta$CH and O (Mg), further implies that our grouping of G1, G2, G3, and G4 would correspond to subpopulations defined with [Na/O] or [Na/Mg] in high-resolution studies. The relations between $\delta CN-\delta CH$ and [Na/O], [Na/Mg] are shown in Figure~\ref{fig_comp}. It shows that our G1, G2, G3 and G4 subpopulations, which are defined based on $\delta CN-\delta CH$, are well matched, respectively, with P1, P2, I1+I2, and E subpopulations defined from the [Na/O] and [Na/Mg] (\citealt{Carr15}). Despite this general agreement, we can see some notable differences in population division between the two studies. As the histograms show, our $\delta CN - \delta CH$ index can more clearly separate G1 (P1) and G2 (P2) than does [Na/O] or [Na/Mg]. It is mainly due to the fact that G1 (P1) and G2 (P2) show a larger difference in $\delta$CN compared to that in Na. On the other hand, G3 (I1+I2) and G4 (E) are more separated in [Na/O] and [Na/Mg] than $\delta CN - \delta CH$, which is because G3 (I1+I2) and G4 (E) show larger differences in O and Mg compared to that in $\delta$CH.

Figure~\ref{fig_cn_cnfe} shows a relation between our $\delta$CN and [CN/Fe]\footnote{While the usual definition of [X/Fe] is $[X/Fe] = \log{[X/Fe]_{star}} - \log{[X/Fe]_{\sun}}$, [CN/Fe] in \citet{Carr18} means the value of [N/Fe] with assuming [C/Fe]=0, where [N/Fe] is derived from CN features in near-IR region (8704{\AA}-8836{\AA}) with synthetic spectra fitting. In other words, [CN/Fe] here does not mean the abundance of CN molecules.} adopted from \citet{Carr18}. It shows that our G1 and G2 are clearly identified by $\delta$CN, whereas they have the same [CN/Fe]. The [CN/Fe] seems to work in separating subpopulations only beyond G2. It implies that, for the earlier subpopulations, $\delta$CN measured from $\sim$3900{\AA} is more sensitive to abundance variations than is [CN/Fe] obtained from near-IR region (see also \citealt{Har03,Pan10}). On the other hand, G3 appears to have a higher value of [CN/Fe] than G4. Since our population division between G3 and G4 is not as efficient as that from high-resolution spectroscopy (see Figure~\ref{fig_comp}), this is probably because our G3 is polluted by G4 stars.

\begin{table*}
	\centering
	\caption{Comparison of subpopulation division among different studies.}
	\renewcommand{\arraystretch}{}
	\begin{tabular}
		{>{\centering}m{.27\linewidth} >{\centering}m{.13\linewidth} >{\centering}m{.12\linewidth} >{\centering}m{.15\linewidth} >{\centering}m{.22\linewidth}}
		\hline\hline
		& Nomenclature & G1 & G2 & G3 \& G4 \tabularnewline
		\cline{2-5}
		This work & \multirow{2}{*}{Properties} & CN-weak & CN-intermediate & CN-strong \tabularnewline
		(Low-resolution spectroscopy; & & (CH-strong) & (CH-strong) & (CH-intermediate/weak) \tabularnewline
		\cline{2-5}
		N = 91) & Fraction & 34\% $\pm$ 6\%& 22\% $\pm$ 5\% & 44\% $\pm$ 7\% \tabularnewline
		\cline{1-5}
		
		& Nomenclature & P1 & P2 & I1, I2, \& E \tabularnewline	
		\cline{2-5}	
		Carretta et al. 2015 & \multirow{2}{*}{Properties} & Na-poor & Na-intermediate & Na-rich \tabularnewline
		(High-resolution spectroscopy; & & (O-rich) & (O-rich) & (O-intermediate/poor) \tabularnewline
		\cline{2-5}
		N = 140)& Fraction & 33\% $\pm$ 5\% & 29\% $\pm$ 5\% & 38\% $\pm$ 5\% \tabularnewline
		\cline{1-5}
		
		Milone et al. 2015 & Nomenclature & B (\& A) & C & D \& E \tabularnewline
		\cline{2-5}
		(HST photometry; N $>$ 1,000) & Fraction & 23.2\% $\pm$ 0.9\% & 26.4\% $\pm$ 1.2\% & 50.4\%  $\pm$ 1.6\% \tabularnewline
		\cline{1-5}
	\end{tabular}
	\label{table_comp}
\end{table*}

\section{Discussion} \label{section:discussion}

In this work, we have measured CN, CH, and HK' indices for the 91 RGB stars in NGC 2808 and confirmed the large spreads in CN and CH. Along the anti-correlated sequence based on CN and CH indices, we have divided 4 subpopulations, G1, G2, G3, and G4. A `jump' in $\delta$HK' index between G1 and G4 has been detected. In addition, we found that $\delta$CN is more efficient in dividing G1 and G2 compared to [Na/Fe]. More detailed aspects of these three points are discussed below.

\subsection{Comparison of population division} \label{subsection:comp_popdiv}
In Section~\ref{section:comp_hrs}, we compared subpopulations of NGC 2808 identified from low-resolution spectroscopy with those defined from high-resolution studies. \citet{Mil15}, on the other hand, compared subpopulations of this GC defined based on \textit{HST UV Legacy Survey} photometric data with those from high-resolution spectroscopy. They defined A, B, C, D, and E subpopulations on the plane of $\Delta$(F275W-F814W) and $\Delta$(F336W-F438W) color indices, which are sensitive to O and N abundances, respectively. From cross-matching, they show that B (including A), C, D, and E subpopulations identified from photometry correspond, respectively, to P1, P2, I1\&I2, and E subpopulations defined from high-resolution spectroscopy (\citealt{Carr15}). Therefore, we can compare subpopulations defined by three different types of observation, low-resolution spectroscopy, high-resolution spectroscopy, and \textit{HST} photometry, which are summarized in Table~\ref{table_comp}\footnote{Direct matching of our data with \textit{HST} photometry was not possible due to the different field-of-view (FOV). FOV of UVIS/WFC3 on \textit{HST} is 2{\arcmin}.7x2{\arcmin}.7 around the cluster center, but that for WFCCD is 25'x25' with the targets mainly located at outer region of the cluster.}.

Table~\ref{table_comp} shows that the population ratios\footnote{For the comparison of population ratios, we re-classified subpopulations of each study into three groups.} from the low- and high-resolution spectroscopic studies are consistent each other within the statistical error. \textit{HST} photometry, however, shows a marginal difference from spectroscopic results. The fraction of the primordial population from spectroscopy (G1; P1) is $\sim$34\%, but that from photometry (B \& A) is $\sim$23\%. On the contrary, the most enriched population group from spectroscopy (G3 \& G4; I1, I2, \& E) is $\sim$40\% of the total population, but that from photometry (D \& E) is $\sim$50\%. This marginal difference in population ratios between spectroscopic and photometric studies would be explained by the difference in the radial distribution among subpopulations. A number of investigations have shown that chemically enriched later generation stars are generally observed to be more centrally concentrated than stars with primordial abundances (\citealt{Bel09,Lar11,Nat11,JP12,Mil12}), despite a few exceptional cases (\citealt{jwlee15,Lim16,Dal18}). In case of NGC 2808, \citet{Simi16} reported that D+E populations (G3 and G4 in this study) are more concentrated to the cluster center than is B+C populations (G1 and G2 in this study) using the \textit{HST} data employed by \citet{Mil15}. Therefore, the population ratio from \citet{Mil15} is biased to overestimate D+E populations compared to population B+C. If we only take the outermost data points in Figure 4 of \citet{Simi16} ($\sim$6{\arcmin}.5 from the center of the cluster), the population ratio between B+C and D+E is about 0.6:0.4, which is in good agreement with the results from spectroscopic studies.

\begin{figure*}
	\centering{\includegraphics[scale=1.05]{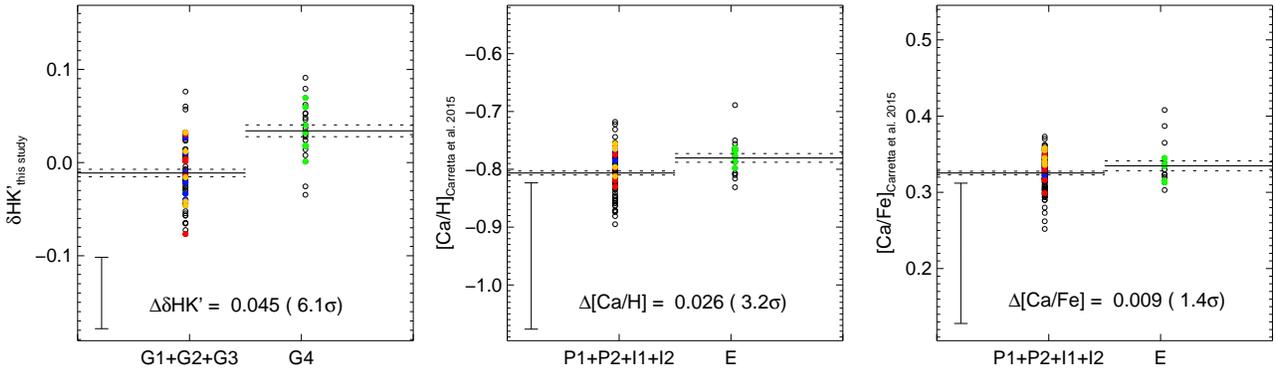}}
	\caption{The relative differences in $\delta$HK' (left), [Ca/H] (middle), and [Ca/Fe] (right) between subpopulations. The color coding and symbols are the same as in Figure~\ref{fig_comp}. The mean value and the error of the mean for each subpopulation are denoted as solid and dashed lines, respectively. Note that G4 (E) is enhanced in both $\delta$HK' and [Ca/H] compared to the combination of the less enriched subpopulations, which are significant at more than 3$\sigma$ level. On the contrary, the difference is negligible in [Ca/Fe].}
	\label{fig_ca}
\end{figure*}

\subsection{Intrinsic Ca spread in NGC 2808} \label{subsection:Ca_spread}
In Section~\ref{section:result_lrs}, we showed a `jump' of Ca II H\&K line strength at G4. A variation in Ca II triplet line strength is also reported by \citet{Hus20}. The spread of Ca line strength in a GC is usually interpreted as an intrinsic star-to-star variation in Ca abundance (\citealt{Sunt80,NF82,Han15,Lim15,Lim17}), a sign of SN II enrichment. Since \citet{Carr15} provides Ca abundances, we can verify whether the G4 (the most enriched population) is indeed enhanced in Ca or not. Therefore, we compared relative differences between the most enriched subpopulation (G4 or E) and the combination of the less enriched subpopulations (G1+G2+G3 or P1+P2+I1+I2) in $\delta$HK', [Ca/H], and [Ca/Fe], because G1 (P1), G2 (P2), and G3 (I1+I2) have indistinguishable $\delta$HK' and [Ca/H]. The left panel of Figure~\ref{fig_ca} clearly shows an enhancement in $\delta$HK' index for G4 compared to the previous generations. A similar trend is observed in the [Ca/H] as shown in the middle panel. The relative difference in [Ca/H] between the subpopulation E and the less enriched subpopulations is 0.026 dex with 3.2$\sigma$ significance. This enhancement in [Ca/H] at the subpopulation E (G4) has been explained by assuming a proton-capture reaction under extreme conditions with low-metallicity, extremely-high temperature (T $>$ 150 MK), and/or increased cross-section (\citealt{Ven12,Carr15,CB21}).
	
Figure~\ref{fig_ca} shows, however, that, unlike [Ca/H], the subpopulation E does not show any statistically significant enhancement in [Ca/Fe] compared to the other subpopulations.\footnote{[Sc/H] and [Sc/Fe] also show a similar trend, although the difference in [Sc/Fe] appears to be statistically significant.} This may indicate that the enhancement in $\delta$HK' or [Ca/H] in the most extreme subpopulation may not be entirely due to the actual enhancement in Ca. One possible explanation for this is the depletion of H caused by significant He enhancement. From the photometric studies of MS and HB stars in NGC 2808, it is already reported that the He mass fraction in the most extreme subpopulation (G4) approaches to $\sim$0.4 (\citealt{Lee05,Pio07,Pas11,Mar14}). This strong He enhancement would lead to the depletion of H in the stellar atmosphere, and therefore, could explain up to 0.08 dex enhancement in [Ca/H]. An extreme Mg depletion can be another candidate for the large enhancement in $\delta$HK' of G4. \citet{Coh10} measured Ca II triplet line strength for RGB stars in NGC 2419, and reported $\sim$0.2 dex spread of Ca abundance in this GC. \citet{Muc12}, however, argued that this spread from Ca II lines is not due to the Ca or Fe abundance spread, but mainly due to the large depletion in Mg abundance. They pointed out the role of Mg in the opacity of cool stars as important electron donors to form the H$^{-}$ ion. A depletion of Mg, therefore, leads to a decrease of the opacity and the electron pressure, which results in stronger Ca II lines at fixed Ca and Fe abundances. Ca and Fe abundances obtained from Ca I and Fe I lines, respectively, indeed do not show any intrinsic spread in this GC (\citealt{Muc12}). The same effect is observed in NGC 5824 (\citealt{DaCosta14,Roe16,Muc18}), another GC showing a large dispersion in Mg abundance. Like NGC 2419 or NGC 5824, NGC 2808 is also one of the GCs which host a strong Mg-depleted subpopulation (\citealt{Carr15,Muc15}), although the spread of Mg in NGC 2808 is relatively small ($\sim$ 0.5 dex) compared to those in NGC 2419 ($\sim$ 2 dex) and NGC 5824 ($\sim$ 1 dex). Therefore, it is possible that a strong Mg depletion can contribute to the `jump' of Ca II H\&K line strength observed at G4 in NGC 2808.

\begin{figure}
	\centering{\includegraphics[scale=0.5]{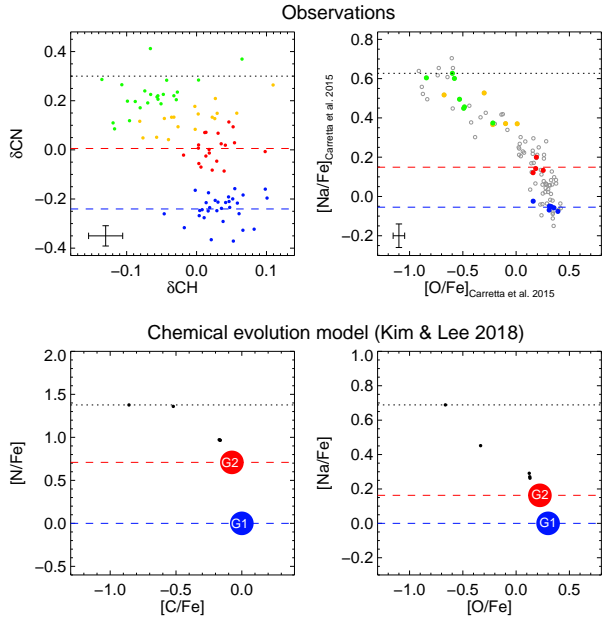}}
	\caption{Comparison of observations with the chemical evolution model of NGC 2808 from \citet{KL18}. The CN and the CH indices are well-known tracers of N and C abundances in GCs. Note that a larger difference in [N/Fe] between G1 and G2 compared to that in [Na/Fe] is naturally predicted in the model.}
	\label{fig_obs_model}
\end{figure}

\subsection{The large difference in CN between G1 and G2} \label{subsection:CN_difference}
Since the CN strength is a well known tracer of N abundance in studies of GCs (see, e.g., \citealt{Smi96}), the large difference in CN strength between G1 and G2 would imply a N enhancement in G2 compared to G1. The large difference in N is also supported by the \textit{HST} photometry for this GC (\citealt{Mil15}). As described in Section~\ref{subsection:comp_popdiv}, \citet{Mil15} compared $\Delta$(F336W-F438W) photometric index, which is sensitive to N abundance, with [Na/Fe] in their Figure 8. The comparison shows that there is a large and discrete difference in $\Delta$(F336W-F438W) between population B (P1) and C (P2), whereas they are overlapped in [Na/Fe]. They estimated that $\Delta$[N/Fe] of population C (P2) with respect to B (P1) is $\sim$0.5 dex, which is calculated from the comparison of observed and synthetic color, and this value is much larger than $\Delta$[Na/Fe] ($\sim$0.2 dex) between them. 

The larger difference in [N/Fe] between G1 and G2 compared to that in [Na/Fe] is also explained in the chemical evolution model of \citet{KL18}, which is developed to reproduce the Na-O anti-correlation generally observed in GCs. In this model, the chemical abundances of later generations are successively enriched by accumulated gas with processed materials ejected by winds of massive stars (WMSs) and AGB stars of previous generations. Since the N yields of WMSs and AGB stars are larger than Na yields, this model naturally predicts that N would be more enhanced than Na in G2 compared to G1 (see Figure~\ref{fig_obs_model}). For subsequent generations (after G3), we do not always find a higher N enhancement than Na when compared to the previous generation. This is probably because abundances of subsequent generations depend more on the adopted star formation history. \citet{Sal20} has recently shown that the initial difference in [N/Fe] between G1 and G2 would be even larger if the effect of the first dredge-up is properly accounted for, while this effect on [Na/Fe] is expected to be negligible (\citealt{KL14}). More detailed quantitative comparison with a chemical evolution model, therefore, should include this effect.

The study of multiple stellar populations with CN band is not restricted to GCs, but can be extended to the Milky Way (MW) bulge. Multiple stellar population model has been suggested to explain the double red clump (RC) phenomenon observed in the CMD of the MW bulge (\citealt{Lee15}). This model predicts chemical abundance differences, such as N and Na, between the two RCs. Since, as shown in this work, CN band strength is more efficient in separating G1 and G2 than is Na abundance, we expect a larger difference between bright and faint RC stars when the CN index is employed as a tracer of multiple populations. Indeed, we have detected a sign of the difference in CN band strength between the stars in the two RC regimes (\citealt{Lee18}). The CN survey for the bulge RC stars in various galactic longitude and latitude fields is in progress (Hong et al. in preparation), and will give a new insight on the structure and formation of the MW bulge.

\begin{acknowledgments}
	We are greatful to the referee for a number of helpful suggestions. We also thank to the staff of LCO for support during the observations. Support for this work was provided by the National Research Foundation of Korea (2017R1A2B3002919). DL gratefully acknowledges funding by the Deutsche Forschungsgemeinschaft (DFG, German Research Foundation) -- Project-ID 138713538 -- SFB 881 (``The Milky Way System'', subproject A03, A05)." S.I.H. acknowledges support provided by Basic Science Research Program through the National Research Foundation of Korea (NRF) funded by the Ministry of Education (Nos. 2018R1D1A1A02085433 and 2020R1I1A1A01052358)
\end{acknowledgments}

\facility{Du Pont (WFCCD)}
\software{IRAF,
	WFCCD Reduction package (\citealt{Pro06,Lim15})}


\appendix
\section{data}
\restartappendixnumbering
\startlongtable
\begin{deluxetable*}{cccccccccccccccc}
	\setlength{\tabcolsep}{0.04in}
	\tabletypesize{\scriptsize}
	\tablewidth{0pt}
	\tablecaption{Index Measurements for the sample stars in NGC 2808\label{tab_data}}
	\tablehead{
		\colhead{ID} & \colhead{RA (deg)} & \colhead{Dec (deg)} & \colhead{$V$ (mag)} & \colhead{S/N} & \colhead{CN} & \colhead{errCN} & \colhead{$\delta$CN} & \colhead{CH} & \colhead{errCH} & \colhead{$\delta$CH} & \colhead{HK'} & \colhead{errHK'} & \colhead{$\delta$HK'} & \colhead{ID$_{Carretta}$}
	}
	\startdata
	N2808-3007 & 137.78148 & -64.80774 & 17.22 & 10.16 & -0.2992 &  0.0915 & -0.1645 &  1.0501 &  0.0424 & -0.0019 &  0.3653 &  0.0772 & -0.0649 & --    \\
	N2808-1002 & 137.78752 & -64.91373 & 14.89 & 34.04 &  0.1735 &  0.0165 &  0.0015 &  1.1558 &  0.0113 &  0.0179 &  0.5872 &  0.0165 & -0.0290 & --    \\
	N2808-1003 & 137.80988 & -64.89328 & 15.29 & 29.51 &  0.2909 &  0.0177 &  0.1712 &  1.0426 &  0.0142 & -0.0807 &  0.6029 &  0.0188 &  0.0184 & 31851 \\
	N2808-1004 & 137.81360 & -64.91374 & 15.67 & 16.39 &  0.2190 &  0.0402 &  0.1493 &  1.1668 &  0.0254 &  0.0575 &  0.5645 &  0.0419 &  0.0103 & --    \\
	N2808-3013 & 137.81650 & -64.86969 & 17.13 &  8.73 & -0.1475 &  0.0800 & -0.0252 &  1.0688 &  0.0418 &  0.0133 &  0.4306 &  0.0711 & -0.0071 & --    \\
	N2808-1006 & 137.82910 & -64.86369 & 15.90 & 17.44 & -0.2779 &  0.0525 & -0.3167 &  1.0831 &  0.0280 & -0.0175 &  0.4699 &  0.0421 & -0.0655 & --    \\
	N2808-3018 & 137.83495 & -64.85916 & 16.26 & 17.39 &  0.2267 &  0.0376 &  0.2350 &  1.0600 &  0.0291 & -0.0275 &  0.5367 &  0.0401 &  0.0299 & --    \\
	N2808-3020 & 137.83864 & -64.83654 & 16.34 & 15.50 &  0.1774 &  0.0311 &  0.1968 &  0.9966 &  0.0233 & -0.0877 &  0.5621 &  0.0315 &  0.0620 & --    \\
	N2808-1007 & 137.84354 & -64.92310 & 15.54 & 27.32 & -0.1304 &  0.0250 & -0.2167 &  1.1723 &  0.0142 &  0.0583 &  0.5428 &  0.0212 & -0.0214 & 30523 \\
	N2808-1008 & 137.84286 & -64.87524 & 14.91 & 19.85 &  0.2523 &  0.0256 &  0.0835 &  1.1351 &  0.0170 & -0.0020 &  0.5984 &  0.0266 & -0.0158 & 32924 \\
	N2808-1009 & 137.84471 & -64.83671 & 15.38 & 21.16 & -0.1068 &  0.0317 & -0.2143 &  1.1614 &  0.0164 &  0.0416 &  0.6041 &  0.0263 &  0.0270 & 35265 \\
	N2808-1010 & 137.85513 & -64.84820 & 15.43 & 19.97 & -0.1372 &  0.0344 & -0.2381 &  1.1536 &  0.0171 &  0.0356 &  0.5832 &  0.0283 &  0.0102 & 34634 \\
	N2808-3026 & 137.86322 & -64.92794 & 16.10 & 17.97 & -0.0696 &  0.0360 & -0.0824 &  1.1157 &  0.0226 &  0.0224 &  0.4896 &  0.0328 & -0.0300 & --    \\
	N2808-1011 & 137.86267 & -64.87892 & 15.73 & 25.56 &  0.1889 &  0.0217 &  0.1277 &  1.0608 &  0.0163 & -0.0461 &  0.5609 &  0.0223 &  0.0120 & 32685 \\
	N2808-1012 & 137.86464 & -64.85864 & 15.17 & 19.76 &  0.4001 &  0.0302 &  0.2650 &  1.0208 &  0.0211 & -0.1069 &  0.6112 &  0.0343 &  0.0174 & 34008 \\
	N2808-3031 & 137.86868 & -64.86993 & 16.26 & 16.41 &  0.2176 &  0.0404 &  0.2258 &  1.0400 &  0.0310 & -0.0475 &  0.5552 &  0.0420 &  0.0482 & --    \\
	N2808-3032 & 137.87207 & -64.94437 & 16.38 & 16.04 & -0.3009 &  0.0827 & -0.2761 &  1.0936 &  0.0398 &  0.0108 &  0.5538 &  0.0622 &  0.0569 & --    \\
	N2808-1014 & 137.88158 & -64.91605 & 14.67 & 20.34 &  0.4213 &  0.0226 &  0.2202 &  1.1137 &  0.0158 & -0.0324 &  0.6658 &  0.0253 &  0.0319 & 30763 \\
	N2808-1015 & 137.88177 & -64.81736 & 15.47 & 20.13 & -0.1190 &  0.0251 & -0.2149 &  1.1303 &  0.0136 &  0.0137 &  0.5594 &  0.0214 & -0.0106 & 13575 \\
	N2808-1016 & 137.88869 & -64.90250 & 15.26 & 24.80 &  0.3391 &  0.0204 &  0.2155 &  1.0678 &  0.0151 & -0.0566 &  0.6464 &  0.0219 &  0.0595 & 31361 \\
	N2808-1017 & 137.89493 & -64.90563 & 15.70 & 15.64 &  0.1160 &  0.0359 &  0.0512 &  1.0508 &  0.0231 & -0.0571 &  0.5393 &  0.0358 & -0.0118 & --    \\
	N2808-1018 & 137.90118 & -64.91302 & 15.14 & 20.78 & -0.1527 &  0.0277 & -0.2914 &  1.1666 &  0.0136 &  0.0379 &  0.5626 &  0.0228 & -0.0334 & 30900 \\
	N2808-1019 & 137.90097 & -64.89154 & 15.71 & 10.75 & -0.2628 &  0.0693 & -0.3271 &  1.1741 &  0.0333 &  0.0663 &  0.5568 &  0.0535 &  0.0060 & --    \\
	N2808-3054 & 137.90620 & -64.91427 & 16.43 & 15.16 & -0.2908 &  0.0502 & -0.2603 &  1.0349 &  0.0267 & -0.0463 &  0.5009 &  0.0389 &  0.0076 & --    \\
	N2808-3061 & 137.91634 & -64.94408 & 16.60 & 16.09 & -0.1387 &  0.0423 & -0.0862 &  1.1144 &  0.0248 &  0.0394 &  0.5063 &  0.0364 &  0.0263 & --    \\
	N2808-3071 & 137.91574 & -64.80714 & 16.01 & 15.29 &  0.1405 &  0.0326 &  0.1163 &  1.0153 &  0.0190 & -0.0812 &  0.4805 &  0.0340 & -0.0461 & 13983 \\
	N2808-1024 & 137.92430 & -64.92281 & 15.95 & 18.44 &  0.0808 &  0.0275 &  0.0487 &  1.0221 &  0.0196 & -0.0767 &  0.5121 &  0.0272 & -0.0192 & --    \\
	N2808-1026 & 137.92705 & -64.88385 & 15.28 & 17.69 & -0.1395 &  0.0374 & -0.2598 &  1.1842 &  0.0172 &  0.0608 &  0.5643 &  0.0310 & -0.0206 & --    \\
	N2808-3087 & 137.93047 & -64.94707 & 16.54 & 17.68 & -0.4162 &  0.0812 & -0.3713 &  1.1298 &  0.0389 &  0.0526 &  0.4277 &  0.0609 & -0.0569 & --    \\
	N2808-1031 & 137.93379 & -64.90678 & 15.69 & 17.73 &  0.2539 &  0.0251 &  0.1874 &  1.0576 &  0.0175 & -0.0508 &  0.6046 &  0.0263 &  0.0525 & --    \\
	N2808-1043 & 137.94591 & -64.81222 & 15.37 & 15.69 & -0.1255 &  0.0408 & -0.2342 &  1.1676 &  0.0203 &  0.0474 &  0.5129 &  0.0354 & -0.0649 & --    \\
	N2808-3129 & 137.94907 & -64.82654 & 17.21 & 10.26 & -0.1149 &  0.0658 &  0.0190 &  1.0934 &  0.0362 &  0.0412 &  0.3761 &  0.0625 & -0.0545 & --    \\
	N2808-3136 & 137.95580 & -64.94306 & 16.78 & 10.96 & -0.2602 &  0.0681 & -0.1838 &  1.1349 &  0.0330 &  0.0665 &  0.4889 &  0.0550 &  0.0234 & --    \\
	N2808-1051 & 137.95392 & -64.81654 & 15.39 & 21.52 &  0.1765 &  0.0196 &  0.0704 &  1.1312 &  0.0125 &  0.0117 &  0.5629 &  0.0200 & -0.0133 &  8198 \\
	N2808-1063 & 137.96939 & -64.91051 & 15.68 & 19.40 & -0.1672 &  0.0307 & -0.2350 &  1.1275 &  0.0168 &  0.0187 &  0.5387 &  0.0255 & -0.0143 & --    \\
	N2808-3168 & 137.96800 & -64.81430 & 17.19 & 11.52 &  0.0899 &  0.0512 &  0.2201 &  0.9980 &  0.0374 & -0.0553 &  0.4605 &  0.0524 &  0.0276 & --    \\
	N2808-3171 & 137.96883 & -64.80650 & 16.82 &  9.14 & -0.0748 &  0.0718 &  0.0070 &  1.0694 &  0.0345 &  0.0026 &  0.5224 &  0.0643 &  0.0602 & --    \\
	N2808-1069 & 137.97200 & -64.82764 & 15.56 & 22.04 &  0.1967 &  0.0189 &  0.1135 &  1.1593 &  0.0127 &  0.0462 &  0.5180 &  0.0201 & -0.0444 & --    \\
	N2808-1070 & 137.97838 & -64.91641 & 15.58 & 22.92 & -0.1101 &  0.0225 & -0.1915 &  1.1583 &  0.0125 &  0.0457 &  0.5755 &  0.0190 &  0.0143 & --    \\
	N2808-1073 & 137.97993 & -64.89509 & 15.59 & 16.47 &  0.1738 &  0.0340 &  0.0940 &  1.0856 &  0.0208 & -0.0265 &  0.5200 &  0.0355 & -0.0403 & --    \\
	N2808-1078 & 137.99362 & -64.88947 & 14.65 & 21.95 & -0.0407 &  0.0248 & -0.2435 &  1.1546 &  0.0138 &  0.0080 &  0.7110 &  0.0202 &  0.0762 & --    \\
	N2808-3216 & 137.99252 & -64.80840 & 16.66 & 11.67 &  0.2229 &  0.0401 &  0.2841 &  1.0754 &  0.0280 &  0.0028 &  0.5077 &  0.0437 &  0.0330 & --    \\
	N2808-3222 & 137.99841 & -64.81222 & 16.38 & 15.98 & -0.2238 &  0.0554 & -0.1995 &  1.1347 &  0.0271 &  0.0517 &  0.5058 &  0.0453 &  0.0087 & --    \\
	N2808-3230 & 138.00664 & -64.91476 & 16.32 & 16.61 & -0.3390 &  0.0578 & -0.3229 &  1.1676 &  0.0255 &  0.0823 &  0.5128 &  0.0440 &  0.0107 & --    \\
	N2808-1087 & 138.01222 & -64.91238 & 15.81 & 20.30 & -0.1617 &  0.0265 & -0.2127 &  1.1346 &  0.0144 &  0.0306 &  0.5032 &  0.0225 & -0.0395 & --    \\
	N2808-3251 & 138.01738 & -64.93042 & 16.34 & 12.52 & -0.3849 &  0.0556 & -0.3656 &  1.1054 &  0.0244 &  0.0210 &  0.4480 &  0.0421 & -0.0522 & --    \\
	N2808-1094 & 138.01527 & -64.83459 & 14.89 & 19.99 &  0.3803 &  0.0246 &  0.2089 &  1.1647 &  0.0155 &  0.0269 &  0.6435 &  0.0269 &  0.0278 & --    \\
	N2808-1096 & 138.02487 & -64.91196 & 14.60 & 30.90 & -0.0376 &  0.0160 & -0.2469 &  1.1530 &  0.0086 &  0.0045 &  0.6632 &  0.0135 &  0.0244 & --    \\
	N2808-3285 & 138.03436 & -64.93041 & 17.33 & 12.69 & -0.0310 &  0.0617 &  0.1187 &  0.9519 &  0.0412 & -0.0959 &  0.4289 &  0.0594 &  0.0078 & --    \\
	N2808-1101 & 138.03734 & -64.93561 & 15.26 & 19.23 &  0.0777 &  0.0251 & -0.0460 &  1.1533 &  0.0150 &  0.0289 &  0.5884 &  0.0237 &  0.0015 & 38244 \\
	N2808-1102 & 138.03697 & -64.92438 & 15.32 & 27.87 &  0.1405 &  0.0159 &  0.0248 &  1.1574 &  0.0101 &  0.0352 &  0.5859 &  0.0157 &  0.0038 & 38967 \\
	N2808-1108 & 138.04079 & -64.80922 & 15.08 & 21.17 &  0.2793 &  0.0246 &  0.1328 &  1.0951 &  0.0162 & -0.0357 &  0.6329 &  0.0256 &  0.0322 &  8826 \\
	N2808-3306 & 138.04482 & -64.81051 & 16.71 & 10.72 & -0.2420 &  0.0595 & -0.1746 &  1.0879 &  0.0298 &  0.0170 &  0.3985 &  0.0510 & -0.0725 & --    \\
	N2808-3371 & 138.07529 & -64.92120 & 16.61 & 11.73 & -0.2510 &  0.0634 & -0.1962 &  1.1752 &  0.0286 &  0.1008 &  0.4610 &  0.0520 & -0.0176 & --    \\
	N2808-3389 & 138.08296 & -64.93560 & 17.31 &  9.72 &  0.1402 &  0.1021 &  0.2865 &  0.9134 &  0.0871 & -0.1354 &  0.4692 &  0.1075 &  0.0461 & --    \\
	N2808-3395 & 138.08127 & -64.80451 & 16.07 & 19.93 &  0.1499 &  0.0286 &  0.1329 &  1.1169 &  0.0198 &  0.0224 &  0.4990 &  0.0300 & -0.0232 & --    \\
	N2808-1155 & 138.08385 & -64.82270 & 15.44 & 19.74 & -0.1477 &  0.0288 & -0.2476 &  1.1468 &  0.0152 &  0.0290 &  0.5323 &  0.0244 & -0.0402 &  7558 \\
	N2808-1158 & 138.08965 & -64.84265 & 15.41 & 22.21 &  0.0825 &  0.0325 & -0.0213 &  1.1003 &  0.0201 & -0.0186 &  0.4981 &  0.0323 & -0.0767 & 54264 \\
	N2808-3413 & 138.09030 & -64.80814 & 17.23 & 11.40 & -0.2956 &  0.0661 & -0.1599 &  1.1500 &  0.0339 &  0.0983 &  0.4254 &  0.0539 & -0.0041 & --    \\
	N2808-1160 & 138.09711 & -64.92525 & 15.78 & 12.18 &  0.3383 &  0.0502 &  0.2834 &  1.0624 &  0.0311 & -0.0427 &  0.5824 &  0.0561 &  0.0372 & --    \\
	N2808-1161 & 138.09665 & -64.85902 & 14.87 & 26.28 &  0.2455 &  0.0203 &  0.0714 &  1.1520 &  0.0129 &  0.0134 &  0.6495 &  0.0206 &  0.0321 & 50561 \\
	N2808-3429 & 138.10388 & -64.91946 & 16.89 &  9.76 & -0.0497 &  0.0663 &  0.0417 &  1.0240 &  0.0360 & -0.0402 &  0.4306 &  0.0627 & -0.0259 & --    \\
	N2808-1162 & 138.10472 & -64.89357 & 15.50 & 25.01 &  0.2825 &  0.0161 &  0.1910 &  1.0456 &  0.0119 & -0.0698 &  0.5684 &  0.0175 &  0.0011 & 42996 \\
	N2808-1165 & 138.10713 & -64.86999 & 15.27 & 21.90 &  0.2543 &  0.0238 &  0.1331 &  1.0913 &  0.0170 & -0.0324 &  0.5435 &  0.0256 & -0.0418 & 48001 \\
	N2808-1167 & 138.11011 & -64.90063 & 15.34 & 20.05 &  0.3765 &  0.0203 &  0.2640 &  1.2307 &  0.0144 &  0.1094 &  0.5233 &  0.0239 & -0.0568 & --    \\
	N2808-3450 & 138.11749 & -64.94714 & 16.84 & 11.77 &  0.0004 &  0.0454 &  0.0855 &  0.9487 &  0.0284 & -0.1172 &  0.5513 &  0.0421 &  0.0911 & --    \\
	N2808-3451 & 138.11632 & -64.90318 & 17.43 & 10.10 & -0.3202 &  0.0772 & -0.1579 &  1.0989 &  0.0377 &  0.0547 &  0.4277 &  0.0619 &  0.0143 & --    \\
	N2808-1170 & 138.12170 & -64.83983 & 15.94 & 19.66 &  0.1609 &  0.0222 &  0.1272 &  1.1252 &  0.0148 &  0.0260 &  0.5337 &  0.0230 &  0.0014 & --    \\
	N2808-1173 & 138.12927 & -64.93564 & 15.03 & 28.16 &  0.0972 &  0.0162 & -0.0557 &  1.1359 &  0.0095 &  0.0032 &  0.6349 &  0.0151 &  0.0303 & 38228 \\
	N2808-1177 & 138.13782 & -64.92680 & 15.86 & 18.39 & -0.1617 &  0.0313 & -0.2056 &  1.1287 &  0.0176 &  0.0266 &  0.5307 &  0.0262 & -0.0078 & --    \\
	N2808-1179 & 138.14699 & -64.90097 & 15.13 & 23.15 &  0.3653 &  0.0212 &  0.2255 &  1.0633 &  0.0162 & -0.0656 &  0.6370 &  0.0232 &  0.0403 & 41688 \\
	N2808-3477 & 138.14740 & -64.81140 & 16.71 & 13.18 & -0.0381 &  0.0844 &  0.0288 &  1.1253 &  0.0473 &  0.0543 &  0.4287 &  0.0820 & -0.0425 & --    \\
	N2808-3478 & 138.15190 & -64.90512 & 17.28 &  9.09 & -0.3912 &  0.1104 & -0.2487 &  1.0664 &  0.0572 &  0.0166 &  0.4088 &  0.0839 & -0.0166 & --    \\
	N2808-1181 & 138.15331 & -64.86250 & 15.23 & 17.37 &  0.3669 &  0.0273 &  0.2398 &  1.0514 &  0.0171 & -0.0740 &  0.6584 &  0.0295 &  0.0694 & 49743 \\
	N2808-3482 & 138.16200 & -64.83875 & 16.69 & 15.80 & -0.2756 &  0.0477 & -0.2102 &  1.1179 &  0.0252 &  0.0465 &  0.4772 &  0.0382 &  0.0051 & --    \\
	N2808-3485 & 138.16679 & -64.81906 & 16.62 & 10.84 &  0.0122 &  0.0571 &  0.0677 &  1.1050 &  0.0310 &  0.0308 &  0.4450 &  0.0572 & -0.0332 & --    \\
	N2808-1184 & 138.18211 & -64.85144 & 15.96 & 20.90 & -0.2373 &  0.0271 & -0.2678 &  1.1021 &  0.0147 &  0.0038 &  0.5088 &  0.0218 & -0.0216 & --    \\
	N2808-1186 & 138.22903 & -64.89172 & 15.54 & 20.20 &  0.2328 &  0.0229 &  0.1460 &  1.1165 &  0.0150 &  0.0024 &  0.5692 &  0.0242 &  0.0047 & --    \\
	N2808-3499 & 138.22774 & -64.82782 & 16.69 & 15.61 &  0.1405 &  0.0370 &  0.2060 &  1.0163 &  0.0269 & -0.0552 &  0.5251 &  0.0374 &  0.0530 & --    \\
	N2808-1188 & 138.23900 & -64.84143 & 15.60 & 21.44 &  0.2548 &  0.0247 &  0.1769 &  1.1091 &  0.0169 & -0.0025 &  0.5683 &  0.0264 &  0.0093 & --    \\
	N2808-3501 & 138.24126 & -64.84245 & 17.42 &  9.73 &  0.2091 &  0.0696 &  0.3694 &  1.1098 &  0.0509 &  0.0650 &  0.4939 &  0.0754 &  0.0793 & --    \\
	N2808-3507 & 138.26326 & -64.91238 & 16.63 & 16.62 &  0.1473 &  0.0344 &  0.2040 &  1.0449 &  0.0251 & -0.0290 &  0.4518 &  0.0370 & -0.0256 & --    \\
	N2808-3508 & 138.26471 & -64.89696 & 17.23 &  9.59 &  0.0623 &  0.0767 &  0.1982 &  0.9528 &  0.0463 & -0.0989 &  0.4535 &  0.0780 &  0.0240 & --    \\
	N2808-1190 & 138.29042 & -64.87961 & 15.23 & 22.33 &  0.1476 &  0.0195 &  0.0200 &  1.1284 &  0.0126 &  0.0029 &  0.5487 &  0.0197 & -0.0406 & --    \\
	N2808-3510 & 138.29019 & -64.83752 & 16.72 & 13.34 & -0.1390 &  0.0502 & -0.0701 &  1.0768 &  0.0315 &  0.0063 &  0.4797 &  0.0439 &  0.0096 & --    \\
	N2808-1192 & 138.31203 & -64.84585 & 15.75 & 15.31 &  0.0458 &  0.0337 & -0.0129 &  1.1244 &  0.0202 &  0.0182 &  0.5599 &  0.0317 &  0.0125 & --    \\
	N2808-3514 & 138.31810 & -64.82907 & 16.34 & 16.21 &  0.3932 &  0.0366 &  0.4122 &  1.0184 &  0.0322 & -0.0660 &  0.5145 &  0.0438 &  0.0142 & --    \\
	N2808-3521 & 138.35162 & -64.87201 & 16.20 & 14.77 &  0.1090 &  0.0377 &  0.1097 &  0.9702 &  0.0276 & -0.1194 &  0.4768 &  0.0387 & -0.0346 & --    \\
	N2808-1193 & 138.37959 & -64.83191 & 15.77 & 15.88 &  0.2050 &  0.0420 &  0.1487 &  1.1350 &  0.0263 &  0.0294 &  0.5496 &  0.0440 &  0.0036 & --    \\
	N2808-3523 & 138.39282 & -64.91596 & 16.67 & 11.37 & -0.0691 &  0.0496 & -0.0070 &  1.1705 &  0.0259 &  0.0981 &  0.4552 &  0.0463 & -0.0190 & --    \\
	N2808-3524 & 138.39613 & -64.90255 & 16.46 & 15.70 &  0.0586 &  0.0381 &  0.0938 &  1.1322 &  0.0240 &  0.0523 &  0.4637 &  0.0382 & -0.0268 & --    
	\enddata
\end{deluxetable*}


\begin{thebibliography}{}
	\bibitem[Bastian et al.(2013)]{Bas13} Bastian, N., Lamers, H.~J.~G.~L.~M., de Mink, S.~E., et al.\ 2013, \mnras, 436, 2398. doi:10.1093/mnras/stt1745
	\bibitem[Bastian \& Lardo(2018)]{BL18} Bastian, N. \& Lardo, C.\ 2018, \araa, 56, 83. doi:10.1146/annurev-astro-081817-051839
	\bibitem[Baumgardt et al.(2008)]{Baum08} Baumgardt, H., Kroupa, P., \& Parmentier, G.\ 2008, \mnras, 384, 1231. doi:10.1111/j.1365-2966.2007.12811.x
	\bibitem[Bellini et al.(2009)]{Bel09} Bellini, A., Piotto, G., Bedin, L.~R., et al.\ 2009, \aap, 493, 959. doi:10.1051/0004-6361:200810880
	\bibitem[Carretta et al.(2003)]{Carr03} Carretta, E., Bragaglia, A., Cacciari, C., et al.\ 2003, \aap, 410, 143. doi:10.1051/0004-6361:20031315
	\bibitem[Carretta et al.(2004)]{Carr04} Carretta, E., Bragaglia, A., \& Cacciari, C.\ 2004, \apjl, 610, L25. doi:10.1086/423034
	\bibitem[Carretta et al.(2006)]{Carr06a} Carretta, E., Bragaglia, A., Gratton, R.~G., et al.\ 2006, \aap, 450, 523. doi:10.1051/0004-6361:20054369
	\bibitem[Carretta(2006)]{Carr06b} Carretta, E.\ 2006, \aj, 131, 1766. doi:10.1086/499565
	\bibitem[Carretta et al.(2009a)]{Carr09a} Carretta, E., Bragaglia, A., Gratton, R.~G., et al.\ 2009, \aap, 505, 117. doi:10.1051/0004-6361/200912096
	\bibitem[Carretta et al.(2009b)]{Carr09b} Carretta, E., Bragaglia, A., Gratton, R., et al.\ 2009, \aap, 505, 139. doi:10.1051/0004-6361/200912097
	\bibitem[Carretta et al.(2010)]{Carr10} Carretta, E., Gratton, R.~G., Lucatello, S., et al.\ 2010, \apjl, 722, L1. doi:10.1088/2041-8205/722/1/L1
	\bibitem[Carretta(2014)]{Carr14} Carretta, E.\ 2014, \apjl, 795, L28. doi:10.1088/2041-8205/795/2/L28
	\bibitem[Carretta(2015)]{Carr15} Carretta, E.\ 2015, \apj, 810, 148. doi:10.1088/0004-637X/810/2/148
	\bibitem[Carretta et al.(2018)]{Carr18} Carretta, E., Bragaglia, A., Lucatello, S., et al.\ 2018, \aap, 615, A17. doi:10.1051/0004-6361/201732324
	\bibitem[Carretta \& Bragaglia(2021)]{CB21} Carretta, E. \& Bragaglia, A.\ 2021, \aap, 646, A9. doi:10.1051/0004-6361/202039392
	\bibitem[Cohen(1999)]{Coh99} Cohen, J.~G.\ 1999, \aj, 117, 2434. doi:10.1086/300860
	\bibitem[Cohen et al.(2010)]{Coh10} Cohen, J.~G., Kirby, E.~N., Simon, J.~D., et al.\ 2010, \apj, 725, 288. doi:10.1088/0004-637X/725/1/288
	\bibitem[D'Antona \& Caloi(2004)]{DC04} D'Antona, F. \& Caloi, V.\ 2004, \apj, 611, 871. doi:10.1086/422334
	\bibitem[D'Antona et al.(2016)]{Dan16} D'Antona, F., Vesperini, E., D'Ercole, A., et al.\ 2016, \mnras, 458, 2122. doi:10.1093/mnras/stw387
	\bibitem[D'Ercole et al.(2010)]{Der10} D'Ercole, A., D'Antona, F., Ventura, P., et al.\ 2010, \mnras, 407, 854. doi:10.1111/j.1365-2966.2010.16996.x
	\bibitem[Dalessandro et al.(2011)]{Dal11} Dalessandro, E., Salaris, M., Ferraro, F.~R., et al.\ 2011, \mnras, 410, 694. doi:10.1111/j.1365-2966.2010.17479.x
	\bibitem[Dalessandro et al.(2018)]{Dal18} Dalessandro, E., Cadelano, M., Vesperini, E., et al.\ 2018, \apj, 859, 15. doi:10.3847/1538-4357/aabb56
	\bibitem[Da Costa et al.(2014)]{DaCosta14} Da Costa, G.~S., Held, E.~V., \& Saviane, I.\ 2014, \mnras, 438, 3507. doi:10.1093/mnras/stt2467
	\bibitem[de Mink et al.(2009)]{DeMink09} de Mink, S.~E., Pols, O.~R., Langer, N., et al.\ 2009, \aap, 507, L1. doi:10.1051/0004-6361/200913205
	\bibitem[Decressin et al.(2007a)]{Dec07a} Decressin, T., Meynet, G., Charbonnel, C., et al.\ 2007, \aap, 464, 1029. doi:10.1051/0004-6361:20066013
	\bibitem[Decressin et al.(2007b)]{Dec07b} Decressin, T., Charbonnel, C., \& Meynet, G.\ 2007, \aap, 475, 859. doi:10.1051/0004-6361:20078425
	\bibitem[Denissenkov \& Hartwick(2014)]{DH14} Denissenkov, P.~A. \& Hartwick, F.~D.~A.\ 2014, \mnras, 437, L21. doi:10.1093/mnrasl/slt133
	\bibitem[Ferraro et al.(1999)]{Fer99} Ferraro, F.~R., Messineo, M., Fusi Pecci, F., et al.\ 1999, \aj, 118, 1738. doi:10.1086/301029
	\bibitem[Gerber et al.(2020)]{Ger20} Gerber, J.~M., Friel, E.~D., \& Vesperini, E.\ 2020, \aj, 159, 50. doi:10.3847/1538-3881/ab607e
	\bibitem[Gieles et al.(2018)]{Gie18} Gieles, M., Charbonnel, C., Krause, M.~G.~H., et al.\ 2018, \mnras, 478, 2461. doi:10.1093/mnras/sty1059
	\bibitem[Gratton et al.(2011)]{Gra11} Gratton, R.~G., Lucatello, S., Carretta, E., et al.\ 2011, \aap, 534, A123. doi:10.1051/0004-6361/201117690
	\bibitem[Gratton et al.(2012)]{Gra12} Gratton, R.~G., Carretta, E., \& Bragaglia, A.\ 2012, \aapr, 20, 50. doi:10.1007/s00159-012-0050-3
	\bibitem[Han et al.(2009)]{Han09} Han, S.-I., Lee, Y.-W., Joo, S.-J., et al.\ 2009, \apjl, 707, L190. doi:10.1088/0004-637X/707/2/L190
	\bibitem[Han et al.(2015)]{Han15} Han, S.-I., Lim, D., Seo, H., et al.\ 2015, \apjl, 813, L43. doi:10.1088/2041-8205/813/2/L43
	\bibitem[Harbeck et al.(2003)]{Har03} Harbeck, D., Smith, G.~H., \& Grebel, E.~K.\ 2003, \aj, 125, 197. doi:10.1086/345570
	\bibitem[Harris(1996; 2010 edition)]{Harris96} Harris, W.~E.\ 1996, \aj, 112, 1487. doi:10.1086/118116
	\bibitem[Husser et al.(2020)]{Hus20} Husser, T.-O., Latour, M., Brinchmann, J., et al.\ 2020, \aap, 635, A114. doi:10.1051/0004-6361/201936508
	\bibitem[Johnson \& Pilachowski(2010)]{JP10} Johnson, C.~I. \& Pilachowski, C.~A.\ 2010, \apj, 722, 1373. doi:10.1088/0004-637X/722/2/1373
	\bibitem[Johnson \& Pilachowski(2012)]{JP12} Johnson, C.~I. \& Pilachowski, C.~A.\ 2012, \apjl, 754, L38. doi:10.1088/2041-8205/754/2/L38
	\bibitem[Karakas \& Lattanzio(2014)]{KL14} Karakas, A.~I. \& Lattanzio, J.~C.\ 2014, \pasa, 31, e030. doi:10.1017/pasa.2014.21
	\bibitem[Kim \& Lee(2018)]{KL18} Kim, J.~J. \& Lee, Y.-W.\ 2018, \apj, 869, 35. doi:10.3847/1538-4357/aaec67
	\bibitem[Krause et al.(2013)]{Kra13} Krause, M., Charbonnel, C., Decressin, T., et al.\ 2013, \aap, 552, A121. doi:10.1051/0004-6361/201220694
	\bibitem[Lardo et al.(2011)]{Lar11} Lardo, C., Bellazzini, M., Pancino, E., et al.\ 2011, \aap, 525, A114. doi:10.1051/0004-6361/201015662
	\bibitem[Latour et al.(2019)]{Lat19} Latour, M., Husser, T.-O., Giesers, B., et al.\ 2019, \aap, 631, A14. doi:10.1051/0004-6361/201936242
	\bibitem[Lee et al.(2009)]{jwlee09} Lee, J.-W., Kang, Y.-W., Lee, J., et al.\ 2009, \nat, 462, 480. doi:10.1038/nature08565
	\bibitem[Lee(2015)]{jwlee15} Lee, J.-W.\ 2015, \apjs, 219, 7. doi:10.1088/0067-0049/219/1/7
	\bibitem[Lee et al.(1999)]{Lee99} Lee, Y.-W., Joo, J.-M., Sohn, Y.-J., et al.\ 1999, \nat, 402, 55. doi:10.1038/46985
	\bibitem[Lee et al.(2005)]{Lee05} Lee, Y.-W., Joo, S.-J., Han, S.-I., et al.\ 2005, \apjl, 621, L57. doi:10.1086/428944
	\bibitem[Lee et al.(2015)]{Lee15} Lee, Y.-W., Joo, S.-J., \& Chung, C.\ 2015, \mnras, 453, 3906. doi:10.1093/mnras/stv1980
	\bibitem[Lee et al.(2018)]{Lee18} Lee, Y.-W., Hong, S., Lim, D., et al.\ 2018, \apjl, 862, L8. doi:10.3847/2041-8213/aad192
	\bibitem[Lim et al.(2015)]{Lim15} Lim, D., Han, S.-I., Lee, Y.-W., et al.\ 2015, \apjs, 216, 19. doi:10.1088/0067-0049/216/1/19
	\bibitem[Lim et al.(2016)]{Lim16} Lim, D., Lee, Y.-W., Pasquato, M., et al.\ 2016, \apj, 832, 99. doi:10.3847/0004-637X/832/2/99
	\bibitem[Lim et al.(2017)]{Lim17} Lim, D., Hong, S., \& Lee, Y.-W.\ 2017, \apj, 844, 14. doi:10.3847/1538-4357/aa79aa
	\bibitem[Marino et al.(2008)]{Mar08} Marino, A.~F., Villanova, S., Piotto, G., et al.\ 2008, \aap, 490, 625. doi:10.1051/0004-6361:200810389
	\bibitem[Marino et al.(2009)]{Mar09} Marino, A.~F., Milone, A.~P., Piotto, G., et al.\ 2009, \aap, 505, 1099. doi:10.1051/0004-6361/200911827
	\bibitem[Milone et al.(2012)]{Mil12} Milone, A.~P., Piotto, G., Bedin, L.~R., et al.\ 2012, \apj, 744, 58. doi:10.1088/0004-637X/744/1/58
	\bibitem[Milone et al.(2017)]{Mil17} Milone, A.~P., Marino, A.~F., Bedin, L.~R., et al.\ 2017, \mnras, 469, 800. doi:10.1093/mnras/stx836
	\bibitem[Marino et al.(2014)]{Mar14} Marino, A.~F., Milone, A.~P., Przybilla, N., et al.\ 2014, \mnras, 437, 1609. doi:10.1093/mnras/stt1993
	\bibitem[Milone et al.(2015)]{Mil15} Milone, A.~P., Marino, A.~F., Piotto, G., et al.\ 2015, \apj, 808, 51. doi:10.1088/0004-637X/808/1/51
	\bibitem[Mucciarelli et al.(2012)]{Muc12} Mucciarelli, A., Bellazzini, M., Ibata, R., et al.\ 2012, \mnras, 426, 2889. doi:10.1111/j.1365-2966.2012.21847.x
	\bibitem[Mucciarelli et al.(2015)]{Muc15} Mucciarelli, A., Bellazzini, M., Merle, T., et al.\ 2015, \apj, 801, 68. doi:10.1088/0004-637X/801/1/68
	\bibitem[Mucciarelli et al.(2018)]{Muc18} Mucciarelli, A., Lapenna, E., Ferraro, F.~R., et al.\ 2018, \apj, 859, 75. doi:10.3847/1538-4357/aaba80
	\bibitem[Nataf et al.(2011)]{Nat11} Nataf, D.~M., Gould, A., Pinsonneault, M.~H., et al.\ 2011, \apj, 736, 94. doi:10.1088/0004-637X/736/2/94
	\bibitem[Norris \& Freeman(1982)]{NF82} Norris, J. \& Freeman, K.~C.\ 1982, \apj, 254, 143. doi:10.1086/159717
	\bibitem[Norris \& Smith(1983)]{NS83} Norris, J. \& Smith, G.~H.\ 1983, \apj, 275, 120. doi:10.1086/161517
	\bibitem[Norris(1987)]{N87} Norris, J.\ 1987, \apjl, 313, L65. doi:10.1086/184832
	\bibitem[Pancino et al.(2010)]{Pan10} Pancino, E., Rejkuba, M., Zoccali, M., et al.\ 2010, \aap, 524, A44. doi:10.1051/0004-6361/201014383
	\bibitem[Pasquini et al.(2011)]{Pas11} Pasquini, L., Mauas, P., K{\"a}ufl, H.~U., et al.\ 2011, \aap, 531, A35. doi:10.1051/0004-6361/201116592
	\bibitem[Piotto et al.(2007)]{Pio07} Piotto, G., Bedin, L.~R., Anderson, J., et al.\ 2007, \apjl, 661, L53. doi:10.1086/518503
	\bibitem[Prochaska et al.(2006)]{Pro06} Prochaska, J.~X., Weiner, B.~J., Chen, H.-W., et al.\ 2006, \apj, 643, 680. doi:10.1086/503184
	\bibitem[Renzini et al.(2015)]{Ren15} Renzini, A., D'Antona, F., Cassisi, S., et al.\ 2015, \mnras, 454, 4197. doi:10.1093/mnras/stv2268
	\bibitem[Roederer et al.(2016)]{Roe16} Roederer, I.~U., Mateo, M., Bailey, J.~I., et al.\ 2016, \mnras, 455, 2417. doi:10.1093/mnras/stv2462
	\bibitem[Salaris et al.(2020)]{Sal20} Salaris, M., Usher, C., Martocchia, S., et al.\ 2020, \mnras, 492, 3459. doi:10.1093/mnras/staa089
	\bibitem[Silich \& Tenorio-Tagle(2017)]{ST17} Silich, S. \& Tenorio-Tagle, G.\ 2017, \mnras, 465, 1375. doi:10.1093/mnras/stw2879
	\bibitem[Simioni et al.(2016)]{Simi16} Simioni, M., Milone, A.~P., Bedin, L.~R., et al.\ 2016, \mnras, 463, 449. doi:10.1093/mnras/stw2003
	\bibitem[Simpson et al.(2017)]{Simp17} Simpson, J.~D., Martell, S.~L., \& Navin, C.~A.\ 2017, \mnras, 465, 1123. doi:10.1093/mnras/stw2781
	\bibitem[Smith et al.(1996)]{Smi96} Smith, G.~H., Shetrone, M.~D., Bell, R.~A., et al.\ 1996, \aj, 112, 1511. doi:10.1086/118119
	\bibitem[Smith et al.(2013)]{Smi13} Smith, G.~H., Modi, P.~N., \& Hamren, K.\ 2013, \pasp, 125, 1287. doi:10.1086/673753
	\bibitem[Smith(2015)]{Smi15} Smith, G.~H.\ 2015, \pasp, 127, 825. doi:10.1086/683012
	\bibitem[Smolinski et al.(2011)]{Smol11} Smolinski, J.~P., Martell, S.~L., Beers, T.~C., et al.\ 2011, \aj, 142, 126. doi:10.1088/0004-6256/142/4/126
	\bibitem[Sneden et al.(1992)]{Sne92} Sneden, C., Kraft, R.~P., Prosser, C.~F., et al.\ 1992, \aj, 104, 2121. doi:10.1086/116388
	\bibitem[Suntzeff(1980)]{Sunt80} Suntzeff, N.~B.\ 1980, \aj, 85, 408. doi:10.1086/112689
	\bibitem[Suntzeff \& Smith(1991)]{SS91} Suntzeff, N.~B. \& Smith, V.~V.\ 1991, \apj, 381, 160. doi:10.1086/170638
	\bibitem[Timmes et al.(1995)]{Timm95} Timmes, F.~X., Woosley, S.~E., \& Weaver, T.~A.\ 1995, \apjs, 98, 617. doi:10.1086/192172
	\bibitem[Ventura et al.(2012)]{Ven12} Ventura, P., D'Antona, F., Di Criscienzo, M., et al.\ 2012, \apjl, 761, L30. doi:10.1088/2041-8205/761/2/L30
	\bibitem[Vollmann \& Eversberg(2006)]{VE06} Vollmann, K. \& Eversberg, T.\ 2006, Astronomische Nachrichten, 327, 862. doi:10.1002/asna.200610645
\end{thebibliography}
\end{document}